\documentclass[12pt]{article}

\renewcommand{\d}{{\rm d}}

\newcommand{\son}{\sigma^1}
\newcommand{\tson}{\tilde{\sigma}^1}

\begin{document}

\title{First order minisuperspace model for the Faddeev formulation of gravity
}

\author{V.M. Khatsymovsky \\
 {\em Budker Institute of Nuclear Physics} \\ {\em of Siberian Branch Russian Academy of Sciences} \\ {\em
 Novosibirsk,
 630090,
 Russia}
\\ {\em E-mail address: khatsym@gmail.com}}
\date{}
\maketitle
\begin{abstract}
Faddeev formulation of general relativity (GR) is considered where the metric is composed of ten vector fields or a ten-dimensional tetrad. Upon partial use of the field equations, this theory results in the usual GR.

Earlier we have proposed some minisuperspace model for the Faddeev formulation where the tetrad fields are piecewise constant on the polytopes like 4-simplices or, say, cuboids into which ${\rm I \hspace{-3pt} R}^4$ can be decomposed.

Now we study some representation of this (discrete) theory, an analogue of the Cartan-Weyl connection-type form of the Hilbert-Einstein action in the usual continuum GR.

\end{abstract}


MSC classes: 83C27; 53C05

keywords: Einstein theory of gravity; composite metric; minisuperspace model; lattice gravity; Faddeev gravity; piecewise flat spacetime; connection

\section{Introduction}

Minisuperspace models may help in studying such an essentially nonlinear theory of gravity as GR. These allow one to work with a countable number of the degrees of freedom. This may be useful in quantum framework as a kind of discretization \cite{Ham'} because of nonrenormalizability of the original continuum GR. The most natural way to get a countable set of the degrees of freedom in gravity, that is, in the curved geometry, is to concentrate ourselves on the metric field distributions $g_{\lambda \mu} (x )$ describing the {\it simplicial complex} or the piecewise flat spacetimes composed of the flat 4D tetrahedra or {\it 4-simplices} \cite{piecewiseflat=simplicial'}. These spacetimes can be chosen arbitrarily close in some sense to any given Riemannian spacetime, and GR on them is known as Regge calculus \cite{Regge'}; see, e. g., review \cite{RegWil'}. The Regge action is proportional to
\begin{equation}                                                            
\label{S-Regge} \sum_{\sigma^2}{\alpha_{\sigma^2}A_{\sigma^2}},
\end{equation}

\noindent where $A_{\sigma^2}$ is the area of the triangle (2-simplex) $\sigma^2$, $\alpha_{\sigma^2}$ is the angle defect on this triangle, summation is over all the 2-simplices $\sigma^2$. The Causal Dynamical Triangulations approach related to the Regge calculus has lead to important results in quantum gravity \cite{cdt}.

There are opportunities to get some other minisuperspace formulation of the theory of gravity starting with an alternative set of variables. One such formulation proposed by Faddeev \cite{Fad} uses $D = 10$ covariant vector fields $f^A_\lambda (x)$ or $D$-dimensional tetrad as the main variables, and the metric is a bilinear function of them,
\begin{eqnarray}\label{metric}                                              
g_{\lambda \mu} = f^A_\lambda f_{\mu A}.
\end{eqnarray}

\noindent Here, the Latin capitals $A, B, \dots = 1, \dots , D$ refer to an Euclidean (or Minkowsky) $D$-dimensional spacetime. To simplify notations, the case of the Euclidean metric signature is considered. A priori, this starting point means a different physical content of the theory compared to GR. It turns out that classically or on the equations of motion it is equivalent to GR. However, in the framework of the quantum approach or "off-shell", we obtain, in general, a different theory.

Faddeev gravity stems from some modification of the so-called embedded theories of gravity \cite{Deser,Paston1,Paston}. Upon choosing $f^A_\lambda = \partial_\lambda f^A$, Faddeev gravity would become a theory of this type, the field variables $f^A$ being coordinates of the four-dimensional hypersurface (our spacetime) embedded into a flat $D$-dimensional spacetime, but genuine Faddeev formulation regards $f^A_\lambda$ as {\it independent} variables.

More generally, the Faddeev gravity has to do with gravity theories where the metric spacetime is not a fundamental physical concept but emerges from a non-spatio-temporal structure present in a more complete theory of interacting fundamental constituents (appearing, e. g., in the context of string theory) \cite{Chiu}.

An important point of the Faddeev approach is introducing a connection $\tilde{\Gamma}_{\lambda \mu\nu} = f^A_\lambda f_{\mu A, \nu} ~~~ (f_{\mu A, \nu} \equiv \partial_\nu f_{\mu A}), ~~~ \tilde{\Gamma}^\lambda_{\mu\nu} = g^{\lambda\rho} \tilde{\Gamma}_{\rho \mu \nu}$ alternative to the unique torsion-free Levi-Civita one, $\Gamma^\lambda_{ \mu\nu}$, and the corresponding curvature tensor $K^\lambda_{\mu \nu \rho}$ instead of the Riemannian one $R^\lambda_{\mu \nu \rho}$. The action takes the form
\begin{equation}\label{Fad action}                                          
\hspace{-10mm} S = \int {\cal L} \d^4 x = \int K^\lambda_{\mu \lambda \rho} g^{\mu \rho} \sqrt {g} \d^4 x = \int \Pi^{AB} (f^\lambda_{A, \lambda} f^\mu_{B, \mu} - f^\lambda_{A, \mu} f^\mu_{B, \lambda}) \sqrt {g} \d^4 x.
\end{equation}

\noindent Here, the projectors onto the {\it vertical} $\Pi_{AB}$ (and {\it horizontal} $\Pi_{||AB}$) directions are
\begin{equation}\label{Pi, Pi||}                                            
\Pi_{AB} = \delta_{AB} - f^\lambda_A f_{\lambda B}, ~~~ \Pi_{||AB} = f^\lambda_A f_{\lambda B}.
\end{equation}

\noindent Varying the action by $\Pi_{AB} \delta / \delta f^\lambda_B$ gives
for the torsion $T^\lambda_{ \mu \nu } = f^{\lambda A} (f_{\mu A , \nu} - f_{\nu A , \mu})$
\begin{equation}\label{V lambda A}                                          
b^\nu{}_{\nu A} T^\mu_{\lambda \mu} + b^\nu{}_{\mu A} T^\mu_{\nu \lambda} + b^\nu{}_{\lambda A} T^\mu_{\mu \nu} = 0.
\end{equation}

\noindent Here, $b^\lambda{}_{\mu A} = \Pi_{AB} f^{\lambda B}_{, \mu}$. The index $A$ of the projected by $\Pi_{AB}$ expression takes on effectively $D - 4$ values. Thus we have $4 (D - 4)$ independent equations forming a linear system for $T^\lambda_{\mu \nu}$, the number sufficient to ensure that $4 \times 6 = 24$ components of $T^\lambda_{\mu \nu}$ be zero \cite{Fad} just at $D \geq 10$. For definiteness, we can take $D = 10$. If $T^\lambda_{\mu \nu} = 0$, then $\tilde{\Gamma}^\lambda_{\mu\nu} = \Gamma^\lambda_{\mu\nu}$, $K^\lambda_{\mu \nu \rho} = R^\lambda_{\mu \nu \rho}$ and the action (\ref{Fad action}) is just the Hilbert-Einstein one.

Faddeev action can be generalized by adding a parity-odd term \cite{Kha0},
\begin{eqnarray}\label{Faddeev}                                             
S = \int \Pi^{AB} \left [ (f^\lambda_{A, \lambda} f^\mu_{B, \mu} - f^\lambda_{A, \mu} f^\mu_{B, \lambda}) \sqrt {g} - \frac{1}{\gamma_{\rm F}} \epsilon^{\lambda \mu \nu \rho} f_{\lambda A, \mu} f_{\nu B, \rho} \right ] \d^4 x.
\end{eqnarray}

\noindent Upon using field equations, this term would result in $\sim \epsilon^{\lambda\mu\nu\rho} R_{\lambda\mu\nu\rho} = 0$ in the GR action. $\gamma_{\rm F}$ is an analog of the Barbero-Immirzi parameter $\gamma$ \cite{Barb,Imm,Holst,Fat} in the Cartan-Weyl form of GR, but survives in the original second order Faddeev action as well.

\begin{figure}[h]
\unitlength 0.7pt
\begin{picture}(180,180)(-200,-90)
\put(0,0){\line(-1,0){100}}
\put(0,0){\line(0,-1){95}}
\put(0,0){\vector(1,0){100}}
\put(0,0){\vector(0,1){95}}

\put(0,0){\circle{44}}
\put(18,14){\vector(-1,1){4}}
\put(83,31){$\sigma^3_1$}
\put(41,82){$\sigma^3_2$}
\put(5,90){$x^2$}
\put(105,-3){$x^1$}
\put(-49,65){$\sigma^3_3$}
\put(-100,37){$\sigma^3_{i-1}$}
\put(-85,-67){$\sigma^3_i$}
\put(-26,-93){$\sigma^3_{i+1}$}
\put(62,-66){$\sigma^3_{n-1}$}
\put(83,-17){$\sigma^3_n$}
\put(4,1){$\sigma^{\! 2}$}
\put(-13,-13){$O$}
\put(19,17){$C$}
\put(40,40){$\sigma^4_2$}
\put(55,7){$\sigma^4_1$}
\put(-15,50){$\sigma^4_3$}
\put(-53,35){\dots}
\put(-60,-15){$\sigma^4_i$}
\put(-40,-50){$\sigma^4_{i+1}$}
\put(10,-50){\dots}
\put(47,-28){$\sigma^4_n$}

\thicklines

\put(17.5,7){\line(5,2){62.5}}
\put(0,0){\line(1,2){40}}
\put(0,0){\line(-2,3){40}}
\put(0,0){\line(-5,2){80}}
\put(-15,-12){\line(-5,-4){55}}
\put(0,0){\line(-1,-5){16}}
\put(0,0){\line(6,-1){80}}
\put(0,0){\line(1,-1){60}}

\end{picture}

\caption{Some neighborhood of a triangle $\sigma^2$ shared by 3- and 4-simplices.}
\label{sigma2}
\end{figure}
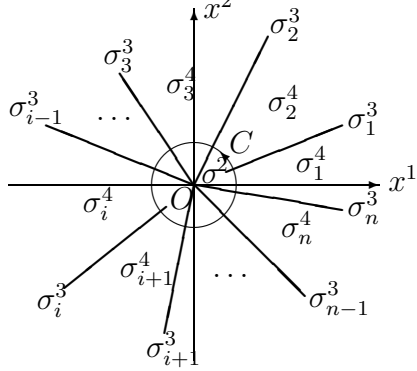

A feature of the Faddeev formulation is finiteness of the action on the discontinuous field configurations because there is no the square of any derivative in the action (\ref{Faddeev}). Discontinuous fields $f^A_\lambda (x)$ mean discontinuous metric $g_{\lambda \mu}$ and thus the possibility to use the simplicial manifold for the minisuperspace where the edge lengths of the different 4-simplices are chosen freely and independently and the different 4-simplices {\it may not coincide} on their common faces. We assume that the fields $f^\lambda_A (x )$ are defined on ${\rm I \hspace{-3pt} R}^4$, the set of points $x = (x^1, x^2, x^3, x^4)$. We can imagine that ${\rm I \hspace{-3pt} R}^4$ is divided (by the hypersurfaces $a_\lambda x^\lambda + b = 0$ or mathematical hyperplanes) into polytopes like the 4-simplices or parallelepipeds and take $f^\lambda_A (x )$ to be constant in each polytope. In figure \ref{sigma2}, the star of a triangle $\sigma^2$ is shown, that is, the set of all the simplices meeting at $\sigma^2$, and the values of $f^\lambda_A (x )$ in the 4-simplices are $f^\lambda_A (\sigma^4_i )$.

In the previous paper \cite{Kha}, we have found the Faddeev action for this minisuperspace. The contribution to the action (\ref{Faddeev}) comes from the 2D faces (triangles), and that from $\sigma^2$ of figure \ref{sigma2} is
\begin{eqnarray}\label{df df Pi V det Dx+gamma}                             
\frac{1}{2} \Pi^{AB} (\sigma^2 ) \sum^n_{i=1} \left \{ \left [ f^{\sigma^1_1}_A ( \sigma^4_i ) f^{\sigma^1_2}_B ( \sigma^4_{i+1} ) - f^{\sigma^1_1}_A ( \sigma^4_{i+1} ) f^{\sigma^1_2}_B ( \sigma^4_i ) \right ] \sqrt{\det \| g_{\sigma^1_\lambda \sigma^1_\mu} \|} \right. \nonumber \\ \left.
- \frac{1}{\gamma_{\rm F}} \left [ f_{\sigma^1_4 A }( \sigma^4_i ) f_{\sigma^1_3 B} ( \sigma^4_{i+1} ) - f_{\sigma^1_4 A } ( \sigma^4_{i+1} ) f_{\sigma^1_3 B} ( \sigma^4_i ) \right ] \right \}
\end{eqnarray}

\noindent where $ g_{\sigma^1_\lambda \sigma^1_\mu} \! = \! f^A_{\sigma^1_\lambda } f_{ \sigma^1_\mu A }$ are metric edge components, $\Pi^A_B = \delta^A_B - \sum_\lambda f^A_{\sigma^1_\lambda } f_B^{\sigma^1_\lambda }$, and $\Pi^{AB} (\sigma^2 )$, $\sqrt{\det \| g_{\sigma^1_\lambda \sigma^1_\mu} \|}$ are some effective values on $\sigma^2$. The co- and contravariant world vector {\it edge} components $f^A_{\sigma^1 }$, $f_A^{\sigma^1 }$ are defined as
\begin{equation}\label{contravar-edge}                                      
f^A_{\sigma^1 } = f^A_\lambda \Delta x^\lambda_{\sigma^1 }, ~~~ f^\lambda_A (\sigma^4_i ) = \sum_\mu f^{\sigma^1_\mu}_A (\sigma^4_i ) \Delta x^\lambda_{\sigma^1_\mu}, ~~~ \Delta x^\lambda_{\sigma^1} = x^\lambda (\sigma^0_2 ) - x^\lambda (\sigma^0_1 )
\end{equation}

\noindent where $\sigma^1_1$, $\sigma^1_2$, $\sigma^1_3$, $\sigma^1_4$ span some $\sigma^4_{i_0} \supset \sigma^2$, and $x^\lambda (\sigma^0_1 )$, $x^\lambda (\sigma^0_2 )$ are the coordinates of the ending vertices of the edge $\sigma^1$.

Some first order action for the Faddeev formulation considered in our paper \cite{Kha0} is the sum of the SO(10) Cartan-Weyl action $S_{SO(10)}$ and certain term $S_\omega$ 
violating the SO(10) local symmetry and linear in Lagrange multipliers $\Lambda^\lambda_{[\mu \nu]} = - \Lambda^\lambda_{[\nu \mu]}$,
\begin{eqnarray}\label{S Imm full}                                              
& & S = S_{SO(10)} + S_\omega, ~~~ S_\omega = \int f^{\lambda A} f^{\mu B} \Lambda^\nu_{[\lambda \mu]} \omega_{\nu AB} \sqrt{g} \d^4 x , \nonumber \\ & & S_{SO(10)} = \int \left ( f^\lambda_A f^\mu_B + \frac{1}{2 \gamma_{\rm F} \sqrt{g}} \epsilon^{\lambda \mu \nu \rho} f_{\nu A} f_{\rho B} \right ) R_{\lambda \mu}^{AB} (\omega ) \sqrt{g} \d^4 x, \\ & & R_{\lambda \mu AB} (\omega ) = \partial_\lambda \omega_{\mu AB} - \partial_\mu \omega_{\lambda AB} + (\omega_\lambda \omega_\mu - \omega_\mu \omega_\lambda)_{AB}. \nonumber
\end{eqnarray}

\noindent That is, the action is $S_{SO(10)}$, and certain constraint (violating the local SO(10)) is imposed,
\begin{equation}\label{wff}                                                
\omega_{\lambda AB} f^A_\mu f^B_\nu = 0.
\end{equation}

\noindent We note in passing that the equations of motion for $\omega_{\lambda AB}$ give
\begin{equation}\label{omega cont}                                         
\omega_{\lambda AB} f^{\mu B} = - \Pi_{AB} \partial_\lambda f^{\mu B}.
\end{equation}

In quantum theory, the parity-odd term entails important consequences for such fundamental concept as area spectrum. As it has been first established in the Loop Quantum Gravity (LQG) approach to GR, the area spectrum is discrete and proportional to $\gamma$, and the discreteness of area is crucial for the black hole physics where it allows to reproduce by statistical methods the Bekenstein-Hawking relation for the black hole entropy \cite{ash}. This spectrum has the form of the sum of some independent "elementary" spectra as if the surface were composed of {\it independent} elementary areas. Both these ingredients, virtual independence of the 2-simplices and availability of an analog of $\gamma$, make the Faddeev gravity approach probably the only known non-LQG candidate to reproduce the discreteness of area spectrum.

Besides that, the first order formalism is most convenient for performing the canonical Hamiltonian analysis required for quantization.

In the case of usual GR, the discrete first order formulations have been widely addressed. The discrete analogs of the connection and curvature were first considered in \cite{Fro}. An application to the discrete Hamiltonian analysis of gravity was discussed in \cite{Ban3}. Regge calculus with the action $\sum A \sin \alpha$ ($\approx \sum A \alpha$ at small $\alpha$) approximately following (again, at small $\alpha$) by excluding connection variables via eqs of motion from a local theory of the Poincar\'{e} group was discussed in \cite{CasDadMag}. A first order form of Regge calculus was considered in \cite{Bar} where the extra independent variables are the interior dihedral angles of a simplex, with conjugate variables the areas of the triangles.

In the present paper, we consider possible (now discrete) first order representation of the minisuperspace second order action (\ref{df df Pi V det Dx+gamma}) similar to the continuum representation (\ref{S Imm full}) of (\ref{Faddeev}). In Section \ref{Cartan-Weyl}, the problem is reduced to that of the SO(10) connection representation of the Regge action, and the latter is found; the exact connection representation of our paper \cite{our2} of the Regge action using the discrete analogs of the connection and curvature \cite{Fro} is applied, it should be only extended to the $SO(10)$ gauge group. In Section \ref{first-order-discrete}, the discrete form of the first order Faddeev action is given; it is checked that excluding SO(10) connection gives the discrete second order Faddeev action (\ref{df df Pi V det Dx+gamma}). In Conclusion, the particular case of the first order discrete Faddeev action is presented for dividing the coordinate $(x^1, x^2, x^3, x^4)$ set ${\rm I \hspace{-3pt} R}^4$ into cuboids. Using flat cubes without restriction on the approximated smooth metric is possible because of the above possibility of the metric discontinuities.

\section{Discrete Cartan-Weyl SO(10) action}\label{Cartan-Weyl}

We do the quite natural assumption that the SO(10) (ie, invariant w. r. t. the local SO(10)) part of the representation of interest (obtained upon separating out possible linear homogeneous in connection part which violates local SO(10)) is a discrete SO(10) part of the first order Faddeev action. In other words, we can bring the action (\ref{S Imm full}) and the constraint (\ref{wff}) to the discrete form separately.

For the action (\ref{S Imm full}), we note that it can be also viewed as a first-order representation of the GR action generalized by introducing the parity-odd term parameterized by the Barbero-Immirzi parameter $\gamma = \gamma_{\rm F}$. The only difference from the genuine Cartan-Weyl action is dimensionality of the orthogonal gauge group, SO(10) instead of SO(4). One can easily show that excluding connection via the equations of motion still gives the Hilbert-Einstein GR action (by, e g, choosing the gauge such that $f^\lambda_A \neq 0$ only at $A = 1, 2 , 3, 4$).

So, finding a discrete version of (\ref{S Imm full}) can be considered as finding a first order representation of the discrete GR (Regge) action with SO(10) gauge group. For that, it is convenient to rewrite the action using the world covariant variables as
\begin{equation}\label{Cartan-SO(10)+gamma}                                
S_{SO(10)} = \int \epsilon^{\lambda \mu \nu \rho} \left ( \frac{1}{4} \epsilon_{ABCD} f^C_\lambda f^D_\mu + \frac{1}{2 \gamma_{\rm F} } f_{\lambda A} f_{\mu B} \right ) R_{\nu \rho}{}^{AB} (\omega ) \d^4 x
\end{equation}

\noindent where we have introduced an analogue of the perfectly antisymmetric fourth rank tensor in the "horizontal" 4-dimensional subspace (the range of the projector $\Pi_{||AB}$ (\ref{Pi, Pi||})), $\epsilon_{ABCD} = ( \det \| f_{\lambda A} f^A_\mu \| )^{1/2} \epsilon^{\lambda \mu \nu \rho} f_{\lambda A} f_{\mu B} f_{\nu C} f_{\rho D}$. In the discrete version, the tensor $\epsilon_{ABCD}$ becomes a function of the 4-simplex on the piecewise constant simplicial ansatz for $f^A_\lambda$. In terms of the edge vectors,
\begin{equation}\label{eps}                                                
\epsilon_{ABCD} (\sigma^4 ) = \frac{\epsilon^{\tson_1 \tson_2 \tson_3 \tson_4} f_{\tson_1 A} f_{\tson_2 B} f_{\tson_3 C} f_{\tson_4 D}}{\sqrt{\det \| f_{\tson_1 A} f^A_{\tson_2} \|}}.
\end{equation}

\noindent Here, $\epsilon^{\tson_1 \tson_2 \tson_3 \tson_4} = \pm 1$ is the parity of the permutation $(\tson_1 \tson_2 \tson_3 \tson_4 )$ of the quadruple of edges $(\son_1 \son_2 \son_3 \son_4 )$ which span the given 4-simplex. The sum over all the permutations is implied.

In the case of SO(4) group, we can use an exact discrete version of the Cartan-Weyl form of the Hilbert-Einstein action \cite{our2} which gives the Regge action if the connection variables are excluded with the help of the equations of motion.

Using $\epsilon_{ABCD} (\sigma^4 )$ (\ref{eps}), the result of \cite{our2} can be extended to SO(10). This gives the SO(10) part of the discrete first order Faddeev action,
\begin{eqnarray}                                                           
& & \hspace{-15mm} S^{\rm discr}_{SO(10)} \! = \! 2 \! \sum_{\sigma^2} \! A(\sigma^2 ) \! \left \{ \! \arcsin \! \left [ \! \frac{v_{\sigma^2 AB }}{2 A(\sigma^2 )} R^{AB}_{\sigma^2} ( \Omega ) \! \right ] \! \! + \! \! \frac{1}{\gamma_{\rm F} } \arcsin \! \left [ \! \frac{V_{\sigma^2 AB }}{2 A(\sigma^2 )} R^{AB}_{\sigma^2} ( \Omega ) \! \right ] \! \right \} \! . 
\end{eqnarray}

\noindent Here, the bivector of the triangle $\sigma^2$ and the dual one are
\begin{equation}                                                           
V^{AB}_{\sigma^2} = \frac{1}{2} ( f^A_{\sigma^1_1} f^B_{\sigma^1_2} - f^B_{\sigma^1_1} f^A_{\sigma^1_2} ), ~~~ v_{\sigma^2 AB} = \frac{1}{2} \epsilon_{ABCD} (\sigma^4 ) V^{CD}_{\sigma^2},
\end{equation}

\noindent the edge vectors $f^A_{\sigma^1_1}$, $f^A_{\sigma^1_2}$ form $\sigma^2$, the area of $\sigma^2$ is $A(\sigma^2 ) = \sqrt{ V^{AB}_{\sigma^2} V_{AB \sigma^2} / 2 }$. The {\it curvature} SO(10) matrix $R^{AB}_{\sigma^2} ( \Omega )$ on the triangles $\sigma^2$ is the product of the {\it connection} SO(10) matrices $\Omega_{\sigma^3}$s for the set of $\sigma^3$s meeting at $\sigma^2$ ordered along a closed path encircling $\sigma^2$ and passing through each of these (and only these) $\sigma^3$s,
\begin{equation}                                                           
R_{\sigma^2} = \prod_{ \{ \sigma^3 : ~ \sigma^3\supset\sigma^2 \} }{\Omega^{\epsilon (\sigma^2, \sigma^3)}_{\sigma^3}},
\end{equation}

\noindent where $\epsilon (\sigma^2, \sigma^3) = \pm 1$ is some sign function. This path begins and ends in a 4-simplex $\sigma^4$. That is, $R^{AB}_{\sigma^2}$ is defined in (the frame of) this simplex. The bivectors $V^{AB}_{\sigma^2}, v^{AB}_{\sigma^2}$ (as well as $f^A_{\sigma^1_1}$, $f^A_{\sigma^1_2}$ on which they depends) are also defined in this simplex, and the same simplex appears as an argument in $\epsilon_{ABCD} (\sigma^4 )$. This 4-simplex $\sigma^4$ is a function of $\sigma^2$: $\sigma^4 = \sigma^4 (\sigma^2 ) \supset \sigma^2$.

To write out the equations of motion for $\Omega_{\sigma^3}$, we add $\sum_{\sigma^3} \mu_{\sigma^3 AB} (\Omega^{CA}_{\sigma^3} \Omega_{\sigma^3 C}{}^B - \delta^{AB} )$, the orthogonality condition for $\Omega_{\sigma^3}$ multiplied by a Lagrange multiplier $\mu_{\sigma^3 AB} = \mu_{\sigma^3 BA}$, to the action and apply the operator $( \Omega_{\sigma^3 C}{}^A \partial / \partial \Omega_{\sigma^3 CB} - \Omega_{\sigma^3 C}{}^B \partial / \partial \Omega_{\sigma^3 CA} )$ to it. The dependence of $R_{\sigma^2}$ on $\Omega_{\sigma^3}$ takes the form $(\Gamma_1 (\sigma^2 , \sigma^3 ) \Omega_{\sigma^3} \Gamma_2 (\sigma^2 , \sigma^3 ))^{\epsilon (\sigma^2, \sigma^3)}$, $\sigma^2 \subset \sigma^3$, $\Gamma_1 (\sigma^2 , \sigma^3 )$, $\Gamma_2 (\sigma^2 , \sigma^3 )$ are SO(10) matrices. The resulting equations read

\begin{eqnarray}\label{dS+gamma/dOmega}                                    
\hspace{-15mm} \sum_{ \hspace{10mm} \{\sigma^2 : ~ \sigma^2 \subset \sigma^3 \} } \hspace{-11mm} \epsilon (\sigma^2, \sigma^3) \Gamma_2 (\sigma^2 , \sigma^3 ) \! \! \left [ \! \frac{v_{\sigma^2 } R_{\sigma^2 } \! + \! R^{\rm T}_{\sigma^2 } v_{\sigma^2 }}{\cos \alpha (\sigma^2 )} + \! \frac{1}{\gamma_{\rm F} } \frac{V_{\sigma^2 } R_{\sigma^2 } \! + \! R^{\rm T}_{\sigma^2 } V_{\sigma^2 }}{\cos \alpha^* (\sigma^2 )} \right ] \! \! \Gamma^{\rm T}_2 (\sigma^2 , \sigma^3 ) \! = \! 0. 
\end{eqnarray}

\noindent Here,
\begin{eqnarray}                                                           
\hspace{-5mm} \alpha (\sigma^2 ) \! = \! \arcsin \! \left [ \! \frac{1}{4} \epsilon_{ABCD} (\sigma^4 ) \frac{V^{AB}_{\sigma^2}}{A(\sigma^2 )} R^{CD}_{\sigma^2} ( \Omega ) \! \right ] \! , \alpha^* (\sigma^2 ) \! = \! \arcsin \! \left [ \! \frac{V_{\sigma^2 AB }}{2 A(\sigma^2 )} R^{AB}_{\sigma^2} ( \Omega ) \! \right ] \! .
\end{eqnarray}

\noindent For the particular ansatz, usual Riemannian (piecewise flat) geometry and $R_{\sigma^2}$ rotating around $\sigma^2$, $\alpha^* = 0$,
\begin{equation}                                                           
v_{\sigma^2 } R_{\sigma^2 } + R^{\rm T}_{\sigma^2 } v_{\sigma^2 } = 2 v_{\sigma^2 } \cos \alpha (\sigma^2 ), ~~~ V_{\sigma^2 } R_{\sigma^2 } + R^{\rm T}_{\sigma^2 } V_{\sigma^2 } = 2 V_{\sigma^2 },
\end{equation}

\noindent (\ref{dS+gamma/dOmega}) reduces to the closure condition for $v_{\sigma^2} + V_{\sigma^2} / \gamma_{\rm F}$, $\sigma^2 \subset \sigma^3$ fulfilled identically, and $S^{\rm discr}_{SO(10)}$ reduces to the Regge action.

\section{The minisuperspace first order Faddeev action}\label{first-order-discrete}

Consider the discrete form of the constraint (\ref{wff}). In this case, the continuum connection $\omega_{\lambda AB}$ there should be replaced\footnote{The form of the discrete version of $\omega_{\lambda AB}$ is not unique: we can choose the antisymmetric part of $\Omega_{\sigma^3 AB}$ mentioned or, say, the generator of $\Omega_{\sigma^3 AB}$. Our actual choice (the antisymmetric part of $\Omega_{\sigma^3 AB}$) is singled out by the simplest functional dependence on $\Omega$ (linear).} by the antisymmetric part of $\Omega_{\sigma^3 AB}$. Therefore, the constraint takes the form
\begin{equation}\label{Omega V}                                            
\Omega_{\sigma^3 AB} V^{AB}_{\sigma^2}{}_{| \sigma^4 (\sigma^3 )} = 0 ~~~ \forall \sigma^2 \subset \sigma^4 (\sigma^3 )
\end{equation}

\noindent (or, equivalently, with $V$ replaced by $v$). The subscript $| \sigma^4 (\sigma^3 )$ means that $V^{AB}_{\sigma^2}$ is defined in $\sigma^4 (\sigma^3 )$ (a function of $\sigma^3$), one of the two 4-simplices sharing $\sigma^3$.

As a result, the discrete version of the full action (\ref{S Imm full}) takes the form
\begin{eqnarray}\label{S-discr}                                            
& & S^{\rm discr} = 2 \sum_{\sigma^2} A(\sigma^2 ) \left \{ \arcsin \left [ \frac{v_{\sigma^2 AB | \sigma^4 (\sigma^2 ) }}{2 A(\sigma^2 )} R^{AB}_{\sigma^2}{}_{ | \sigma^4 (\sigma^2 ) } ( \Omega ) \right ] \right. \nonumber \\ & & + \left. \frac{1}{\gamma_{\rm F} } \arcsin \left [ \frac{V_{\sigma^2 AB | \sigma^4 (\sigma^2 ) }}{2 A(\sigma^2 )} R^{AB}_{\sigma^2}{}_{ | \sigma^4 (\sigma^2 ) } ( \Omega ) \right ] \right \} \nonumber \\ & & + \sum_{\sigma^3} \sum_{\{ \sigma^2 : ~ \sigma^2 \subset \sigma^4 (\sigma^3 ) \} } \Lambda (\sigma^2, \sigma^3 ) \Omega^{AB}_{\sigma^3} v_{\sigma^2 AB | \sigma^4 (\sigma^3)}.
\end{eqnarray}

\noindent Here, $\Lambda (\sigma^2, \sigma^3 )$ are Lagrange multipliers. It is sufficient to put $\Lambda (\sigma^2, \sigma^3 ) \neq 0$ only for any six independent bivectors $v_{\sigma^2}$ in $\sigma^4 (\sigma^3 )$.

Now the equations for $\Omega_{\sigma^3}$ can be regarded as differing from (\ref{dS+gamma/dOmega}) by the presence of some nonzero RHS, $- \sum_{\{ \sigma^2 : ~ \sigma^2 \subset \sigma^4 (\sigma^3 ) \} } \Lambda (\sigma^2, \sigma^3 ) [v_{\sigma^2 } \Omega_{\sigma^3} + \Omega^{\rm T}_{\sigma^3} v_{\sigma^2 }]^{AB}$ and reduce not to the closure condition for bivectors but (upon excluding $\Lambda$) to a weakened form of the latter. Let $\delta f$ be a typical variation of $f^\lambda_A$ when passing from simplex to simplex. We can make sure that we can disregard the $\Lambda$-part if we consider leading orders of magnitude, in particular, $O(\delta f)$ for $\Omega_{\sigma^3} - 1$.

With this observation, we can check that excluding the connection variables from $S^{\rm discr}$ results in the minisuperspace second order Faddeev action if $\delta f$ is taken arbitrarily small, that is, in the continuum limit. First, we take a naive discrete analogue of the continuum connection (\ref{omega cont}),
\begin{equation}\label{omega disc}                                         
\omega_{\sigma^3_j AB} f^{\mu B} (\sigma^4_j ) = \Pi_{AB} (\sigma^4_j ) [f^{\mu B} (\sigma^4_j ) - f^{\mu B} (\sigma^4_{j+1} ) ] + O((\delta f)^2 ),
\end{equation}

\noindent where $\omega_{\sigma^3} = - \omega_{\sigma^3}^{\rm T}$ is the generator of $\Omega_{\sigma^3} = \exp \omega_{\sigma^3}$, and notations correspond to fig. \ref{sigma2}. We check that this indeed solves the equations for $\Omega_{\sigma^3}$. The 2-simplex $\sigma^2$ of fig. \ref{sigma2} has the bivectors
\begin{equation}                                                           
V^{AB} = \frac{1}{2} (f^A_3 f^B_4 - f^A_4 f^B_3), ~~~ v_{AB} = \frac{1}{2} (f^1_A f^2_B - f^2_A f^1_B )\sqrt{g},
\end{equation}

\noindent and the curvature matrix on it should be expanded up to bilinear in $\omega_{\sigma^3}$s terms,
\begin{equation}                                                           
R = \Omega_1 \dots \Omega_i \dots \Omega_n, ~~~ \frac{1}{2} (R - R^{\rm T}) = \sum^n_{i=1} \omega_i + \frac{1}{2} \sum^n_{i < j} (\omega_i \omega_j - \omega_j \omega_i ) + O((\delta f)^3),
\end{equation}

\noindent where $\Omega_j \equiv \Omega_{\sigma^3_j}$, $\omega_j \equiv \omega_{\sigma^3_j}$. Substitute this into the action $S^{\rm discr}$, vary over $\omega_i$, check that (\ref{omega disc}) satisfies the resulting equation for $\omega$, then substitute (\ref{omega disc}) into $S^{\rm discr}$. Thereby the second order action (\ref{df df Pi V det Dx+gamma}) is reproduced where we should replace the edge component indices $\sigma^1_i$ simply by $i$, $i$ = 1, 2, 3, 4.

\section{Conclusion}

We have found the first order representation of the discrete (minisuperspace) Faddeev action (\ref{S-discr}). One way of obtaining the discrete Faddeev action is direct evaluation of the genuine continuum Faddeev action on the piecewise constant distributions of the ten-dimensional tetrad. Thereby, the second order discrete Faddeev action (\ref{df df Pi V det Dx+gamma}) follows. Another way is just to take the continuum first order Faddeev formalism and rewrite it in a discrete form converting basic definitions for the discrete language. This leads to the discrete first order Faddeev action (\ref{S-discr}). Remarkable is that these two objects obtained in the seemingly different ways are in fact related, and (\ref{df df Pi V det Dx+gamma}) follows by excluding the connection type variables from (\ref{S-discr}).

Finally, consider perhaps the simplest minisuperspace ansatz possible because of the possibility of the metric discontinuities. Namely, it seems very simple to divide ${\rm I \hspace{-3pt} R}^4$ into the cuboids
\begin{equation}                                                           
n^\lambda < x^\lambda < n^\lambda + 1, ~~~ \lambda = 1, 2, 3, 4,
\end{equation}

\noindent where $n^1, n^2, n^3, n^4$ are integers. The connection SO(10) matrix $\Omega_\lambda$ is a function of $n^1, n^2, n^3, n^4 $, and acts from the hypercube at $n^\lambda - 1 < x^\lambda < n^\lambda$ to the hypercube at $n^\lambda < x^\lambda < n^\lambda + 1$. Also introduce the operator $T_\lambda$ which shifts the argument $ x^\lambda = n^\lambda $ of any function on the hypercubic lattice by +1, that is, from any site (vertex) to the neighboring site along $x^\lambda$. Our general result (\ref{S-discr}) takes the form
\begin{eqnarray}\label{S-hypercube}                                        
& & \hspace{0mm} S^{\rm discr} = \sum_{\rm sites} \sum_{\lambda, \mu, \nu} \Lambda^\lambda_{[\mu \nu]} \Omega^{AB}_\lambda f^\mu_A f^\nu_B + \sum_{\rm sites} \sum_{\lambda, \mu} \left\{ \frac{\sqrt{ (f^\lambda )^2 (f^\mu )^2 - (f^\lambda f^\mu )^2}}{ \sqrt{\det \| f^\nu f^\rho \|}} \right. \\ & & \phantom{S^{\rm discr} =} \cdot \arcsin \left\{ \frac{f^\lambda_A f^\mu_B - f^\mu_A f^\lambda_B}{2 \sqrt{ (f^\lambda )^2 (f^\mu )^2 - (f^\lambda f^\mu )^2}} \left[ \Omega^{\rm T}_\lambda (T^{\rm T}_\lambda \Omega^{\rm T}_\mu) (T^{\rm T}_\mu \Omega_\lambda) \Omega_\mu \right]^{AB} \right\} + \nonumber \\ & &  \hspace{-7mm} \left. \frac{1}{\gamma_{\rm F}} \sqrt{ \frac{(\epsilon^{\lambda \mu \nu \rho} f_{\nu A} f_{\rho B} )^2}{2}} \arcsin \left\{ \frac{\epsilon^{\lambda \mu \nu \rho} f_{\nu A} f_{\rho B} }{ \sqrt{2 (\epsilon^{\lambda \mu \nu \rho} f_{\nu A} f_{\rho B}  )^2}} \left[ \Omega^{\rm T}_\lambda (T^{\rm T}_\lambda \Omega^{\rm T}_\mu) (T^{\rm T}_\mu \Omega_\lambda) \Omega_\mu \right]^{AB} \right\} \right\} \nonumber
\end{eqnarray}

\noindent (note that $[(f^\lambda )^2 (f^\mu )^2 - (f^\lambda f^\mu )^2] (\det \| f^\nu f^\rho \|)^{-1} \equiv \frac{1}{2} (\epsilon^{\lambda \mu \nu \rho} f_{\nu A} f_{\rho B} )^2$, quadrangle area squared). Here, $\Lambda^\lambda_{[\mu \nu]} = - \Lambda^\lambda_{[\nu \mu]}$ and one can take either $f^\lambda_A$ or $f^A_\lambda$ as the independent tetrad variables. It looks like the sum over plaquettes (here denoted by the pair $\lambda, \mu$) in the Wilson's discrete action in QCD \cite{Wil}. It has relatively simple explicit form, at the same time being (a representation of) a minisuperspace action.

\section*{Acknowledgments}

The present work was supported by the Ministry of Education and Science of the Russian Federation.

\end{document}